\documentclass[a4paper,preprint,byrevtex,showpacs,aip]{revtex4-1}

\DeclareSymbolFont{lettersA}{U}{pxmia}{m}{it}
\DeclareMathAlphabet{\mathsfsl}{OT1}{cmss}{m}{sl}

\SetSymbolFont{lettersA}{bold}{U}{pxmia}{bx}{it}
\DeclareFontSubstitution{U}{pxmia}{m}{it}
\DeclareSymbolFontAlphabet{\mathfrak}{lettersA}
\DeclareMathSymbol{\piup}{\mathord}{lettersA}{"19}
\DeclareMathSymbol{\iTheta}{\mathalpha}{letters}{2}

\usepackage{amsmath}
\usepackage{amssymb}
\usepackage{graphicx}
\usepackage{epstopdf}
\usepackage{bbm}
\usepackage{mathrsfs}
\usepackage{xcolor}
\usepackage[colorlinks=true,pdfstartview=FitV,linkcolor=blue,citecolor=blue,urlcolor=blue]{hyperref}

\makeatletter

\newcommand{\Rmnum}[1]{\expandafter\@slowromancap\romannumeral #1@}
\makeatother

\renewcommand{\vec}[1]{\boldsymbol{#1}}

\newcommand{\ii}{\mathrm{i}}

\newcommand{\blue}[1]{\textcolor{blue}{#1}}

\begin{document}

%%%%%%%%%%%%%
%%% TITLE %%%
%%%%%%%%%%%%%

\title{Loss enhanced transmission and collimation in anisotropic epsilon-near-zero metamaterials}

\author{Lei Sun}
\address{
Department of Mechanical and Aerospace Engineering, \\
Missouri University of Science and Technology, \\
Rolla, MO 65409, USA
}

\author{Simin Feng}
\address{
Physics Division,
Naval Air Warfare Center, \\
China Lake, California 93555, USA
}

\author{Xiaodong Yang}
\email{yangxia@mst.edu}
\address{
Department of Mechanical and Aerospace Engineering, \\
Missouri University of Science and Technology, \\
Rolla, MO 65409, USA
}

\begin{abstract}
We verify the extraordinary transmission enhancement and collimation induced
by the material loss in anisotropic near-zero permittivity (ENZ) metamaterials,
and reveal the physical mechanism of this exotic electromagnetic phenomenon
via the iso-frequency contour (IFC) analysis.
In addition, we demonstrate the possibility in realization of such loss
enhanced transmission of Gaussian beam in realistic silver-germanium multilayered
structures by applying full-wave numerical simulations.
\end{abstract}

%\pacs{TO BE DONE...}
%\ocis{160.1190, 160.3918, 160.4236, 310.4165}

\maketitle

%%%%%%%%%%%%%%%%%
%%% MAIN TEXT %%%
%%%%%%%%%%%%%%%%%

%%%%%% INTRODUCTION %%%%%%%%%%%%%%%%%%%%%%%%%%%%%%%%%%%%%%%%%%%%%%%%%%%%%%%%%
Recently, metamaterials with near-zero permittivity (epsilon-near-zero, ENZ)
emerge into the focus of the extensive explorations in both theory and engineering
due to their anomalous electromagnetic features at microwave and optical frequencies.
ENZ metamaterials have been widely used in lots of exciting applications,
such as directive emission \cite{Enoch2002PRL},
electromagnetic energy squeezing and tunneling
\cite{Silveirinha2006PRL,Silveirinha2007PRB,Liu2008PRL,Edwards2008PRL},
electromagnetic wave phase front shaping \cite{Alu2007PRB,Cheng2012PRL},
electromagnetic transparency \cite{Alu2005PRE,Rainwater2012NJP},
and invisible cloaking \cite{Pendry2006SCI,Schurig2006SCI}.
In the previous work, it was shown theoretically that
in ENZ media loss can induce transparency,
omni-directional collimation, and transmission enhancement
due to the interplay between ENZ and material loss \cite{Feng2012PRL}.
Anisotropic loss can even improve the propagation of oblique incident beams \cite{Feng2012PRL}.
In this Letter, we verify such abnormal transmission enhancement in anisotropic
ENZ metamaterials with full-wave numerical simulations together with the iso-frequency
contour (IFC) analysis, and demonstrate the possibility in experimental realization
of this phenomenon by using silver-germanium multilayered structures.

\begin{figure}[htbp]
    \centering
    \includegraphics[width=13.5cm]{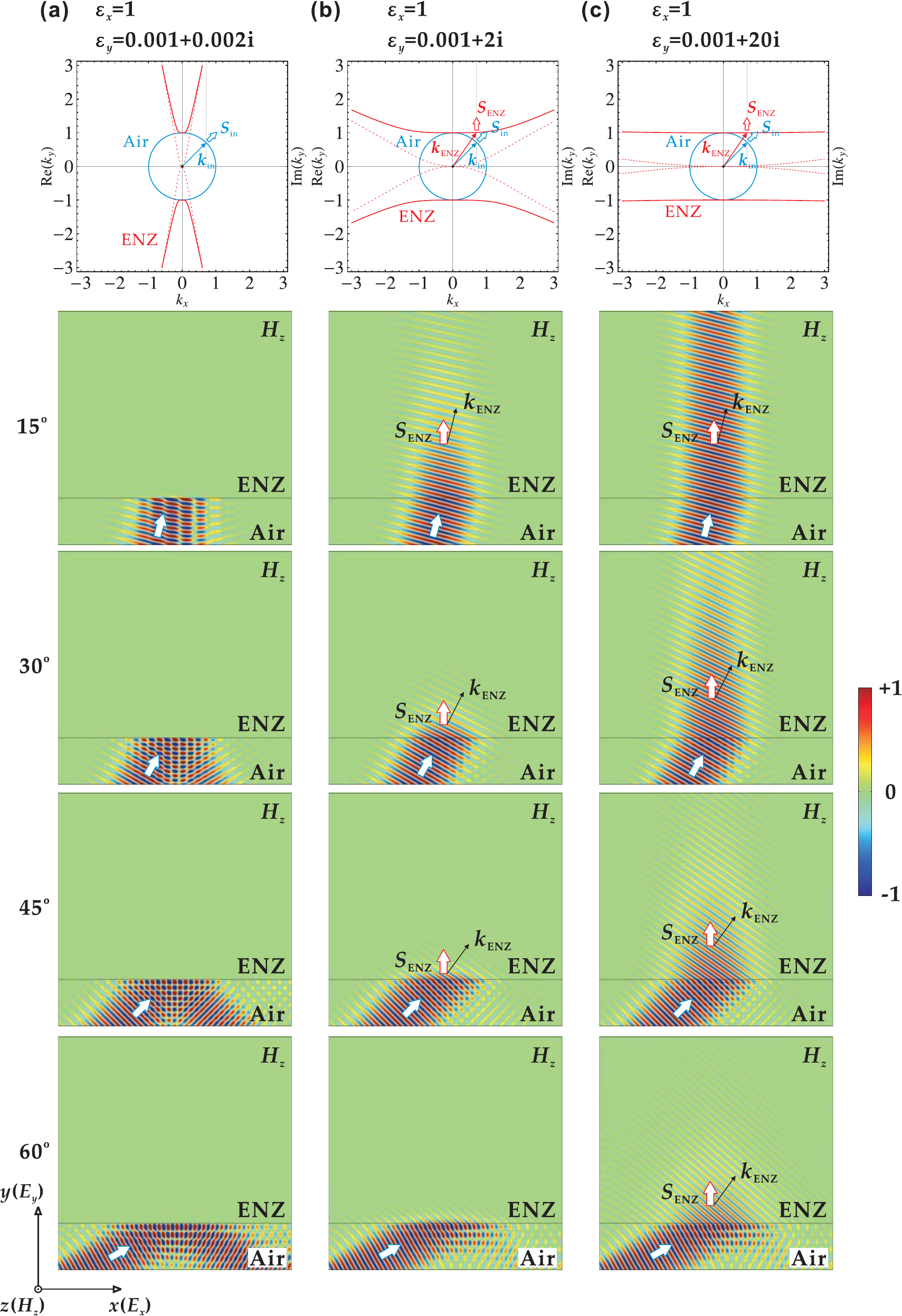}
    \caption{Lei Sun, Simin Feng, and Xiaodong Yang}
    \label{fig:fig1}
\end{figure}

%%%%%% THEORY ANALYSIS %%%%%%%%%%%%%%%%%%%%%%%%%%%%%%%%%%%%%%%%%%%%%%%%%%%%%%
The propagation of a plane electromagnetic wave incident upon the flat interface
(along the $x$-direction) between air and an anisotropic ENZ metamaterial is considered
in two-dimensional space.
The permittivity of air is set to be unity, and the principle components of
the permittivity tensor for the anisotropic ENZ metamaterial are denoted as
$\varepsilon_{x}$ and $\varepsilon_{y}$, respectively.
The permittivity $\varepsilon_{x}$ is also set to be unity, i.e., $\varepsilon_{x}=1$,
while the permittivity $\varepsilon_{y}$
reads $\varepsilon_{y}=\mathrm{Re}(\varepsilon_{y})+\ii\mathrm{Im}(\varepsilon_{y})$
with a near-zero real part $\mathrm{Re}(\varepsilon_{y})$ and a non-zero imaginary part
$\mathrm{Im}(\varepsilon_{y})$ (the material loss).
Regarding a plane electromagnetic wave in TM polarization
(with non-zero field components $E_{x}$, $E_{y}$, and $H_{z}$),
the dispersion relation of the anisotropic ENZ metamaterial reads
\begin{equation}
\label{eq:dispersion}
    \frac{k_{x}^{2}}{\varepsilon_{y}}
	   +\frac{k_{y}^{2}}{\varepsilon_{x}}=k_{0}^{2},
\end{equation}
in which $k_{0}$ is the wave vector in free space.
The wave vector component $k_{x}$ is real and determined by the incident wave,
while $k_{y}$ is complex and it is denoted as $k_{y}=\mathrm{Re}(k_{y})+\ii\mathrm{Im}(k_{y})$,
with the imaginary part $\mathrm{Im}(k_{y})$ representing the propagation loss.
The iso-frequency contours (IFCs) of the anisotropic ENZ metamaterial
can be plotted based on the dispersion relation in Eq.~\eqref{eq:dispersion}.
Figure~\ref{fig:fig1} displays three different IFCs at the frequency of $193.4\,\mathrm{THz}$
($1.55\,\mu\mathrm{m}$) corresponding to different values of permittivity $\varepsilon_{y}$,
which has the same real part of $\mathrm{Re}(\varepsilon_{y})=0.001$,
but different material losses: the low loss $\mathrm{Im}(\varepsilon_{y})=0.002$ (Fig.~\ref{fig:fig1}\blue{(a)}),
the moderate loss $\mathrm{Im}(\varepsilon_{y})=2$ (Fig.~\ref{fig:fig1}\blue{(b)}),
and the high loss $\mathrm{Im}(\varepsilon_{y})=20$ (Fig.~\ref{fig:fig1}\blue{(c)}), respectively.
The IFC of $k_{x}\sim\mathrm{Re}(k_{y})$ is plotted as red solid curve,
while the IFC of $k_{x}\sim\mathrm{Im}(k_{y})$ is plotted as red dashed curve.
The circular IFC of air with radius $k_{0}$ is also plotted as blue solid curve for comparison.
Additionally, the incoming electromagnetic wave in air is represented
by the wave vector $\vec{k}_{\mathrm{in}}$ and the Poynting vector $\vec{S}_{\mathrm{in}}$,
while the electromagnetic wave inside the anisotropic ENZ metamaterial
is represented by the wave vector $\vec{k}_{\scriptscriptstyle\mathrm{ENZ}}$
and the Poynting vector $\vec{S}_{\scriptscriptstyle\mathrm{ENZ}}$.
As a particular example, $\vec{k}_{\mathrm{in}}$ and $\vec{S}_{\mathrm{in}}$
are marked as the incoming electromagnetic wave in air has an incident angle of $45^{\circ}$.

%%%%%% ANALYSIS %%%%%%%%%%%%%%%%%%%%%%%%%%%%%%%%%%%%%%%%%%%%%%%%%%%%%%%%%%
It is clear that for the low loss $\mathrm{Im}(\varepsilon_{y})=0.002$ in Fig.~\ref{fig:fig1}\blue{(a)},
the IFC of $k_{x}\sim\mathrm{Re}(k_{y})$ forms a hyperbolic-like shape,
with a very narrow opening toward the $k_{y}$-direction.
Therefore, the direction of the transmitted power inside the ENZ varies from
normal to nearly parallel to the interface.
Moreover, the high propagating loss, represented by the IFC of $k_{x}\sim\mathrm{Im}(k_{y})$
prevents the electromagnetic wave from transmitting inside the anisotropic
ENZ metamaterial even at small incident angles.
Finite element method (FEM) full-wave simulations are used to illustrate
the wave propagation behavior, where a TM polarized Gaussian beam is applied
to mimic the experimental condition instead of the ideal plane electromagnetic wave.
The Gaussian beam is represented by the distribution of the $H_{z}$ component.
Four different angles of incidence are calculated, including $15^{\circ}$, $30^{\circ}$,
$30^{\circ}$, and $45^{\circ}$.
The simulation results clearly indicate that for all these angle of incidence,
the incoming Gaussian beam cannot couple into and transmit in the anisotropic
ENZ metamaterial, but reflect back at the interface, which is consistent with
the above IFC analysis.
Thus, a low loss of the anisotropic ENZ metamaterial along the $y$-direction
will prevent the transmission due to impedance mismatch.
%%------%%
As the material loss is increased to the moderate loss
$\mathrm{Im}(\varepsilon_{y})=2$ in Fig.~\ref{fig:fig1}\blue{(b)}, the opening of the IFC of
$k_{x}\sim\mathrm{Re}(k_{y})$ toward the $k_{y}$-direction becomes wider,
leading to a flatter curvature.
Besides, the corresponding propagating loss for the oblique incidence also decreases due to the
increased material loss \cite{Feng2012PRL}.
Additionally, because of the flatter curvature of the IFC of $k_{x}\sim\mathrm{Re}(k_{y})$,
the incoming electromagnetic wave with small angle of incidence, e.g., less than $45^{\circ}$,
will transmit inside the anisotropic ENZ metamaterial with the direction
nearly normal to the interface, since the Poynting vector of the electromagnetic wave
should be perpendicular to the IFC of $k_{x}\sim\mathrm{Re}(k_{y})$.
In the simulations, the Gaussian beam can transmit into the anisotropic
ENZ metamaterial with a wide range of the angles of incidence from $15^{\circ}$ to $60^{\circ}$.
However, it is worth noting that the propagation distance of the Gaussian beam
inside the anisotropic ENZ metamaterial is limited by its propagation loss,
especially for the Gaussian beam with large angle of incidence,
which is indicated in the IFC.
Furthermore, the simulations also show that the wave propagation direction
(indicated by the Poynting vector $\vec{S}_{\scriptscriptstyle\mathrm{ENZ}}$)
of the Gaussian beam inside the anisotropic ENZ metamaterial is different from
the direction of its phase velocity (indicated by the wave vector
$\vec{k}_{\scriptscriptstyle\mathrm{ENZ}}$), due to the anisotropic permittivity profile.
%%------%%
For the high loss $\mathrm{Im}(\varepsilon_{y})=20$ in Fig.~\ref{fig:fig1}\blue{(c)},
the IFC of $k_{x}\sim\mathrm{Re}(k_{y})$ becomes almost flat straight lines,
which allows the electromagnetic power inside the ENZ to propagate in the normal
direction for a wide range of the incidence angle,
for example, the Gaussian beam with $60^{\circ}$ angle of incidence
shown in the simulations.
The relatively low propagation loss permits a long propagation distance
of the transmitted electromagnetic wave inside the anisotropic ENZ metamaterial.
As a consequence, the transmission is enabled and enhanced by
the high loss of the anisotropic ENZ metamaterial along the $y$-direction.

\begin{figure}[htbp]
    \centering
    \includegraphics[width=13.5cm]{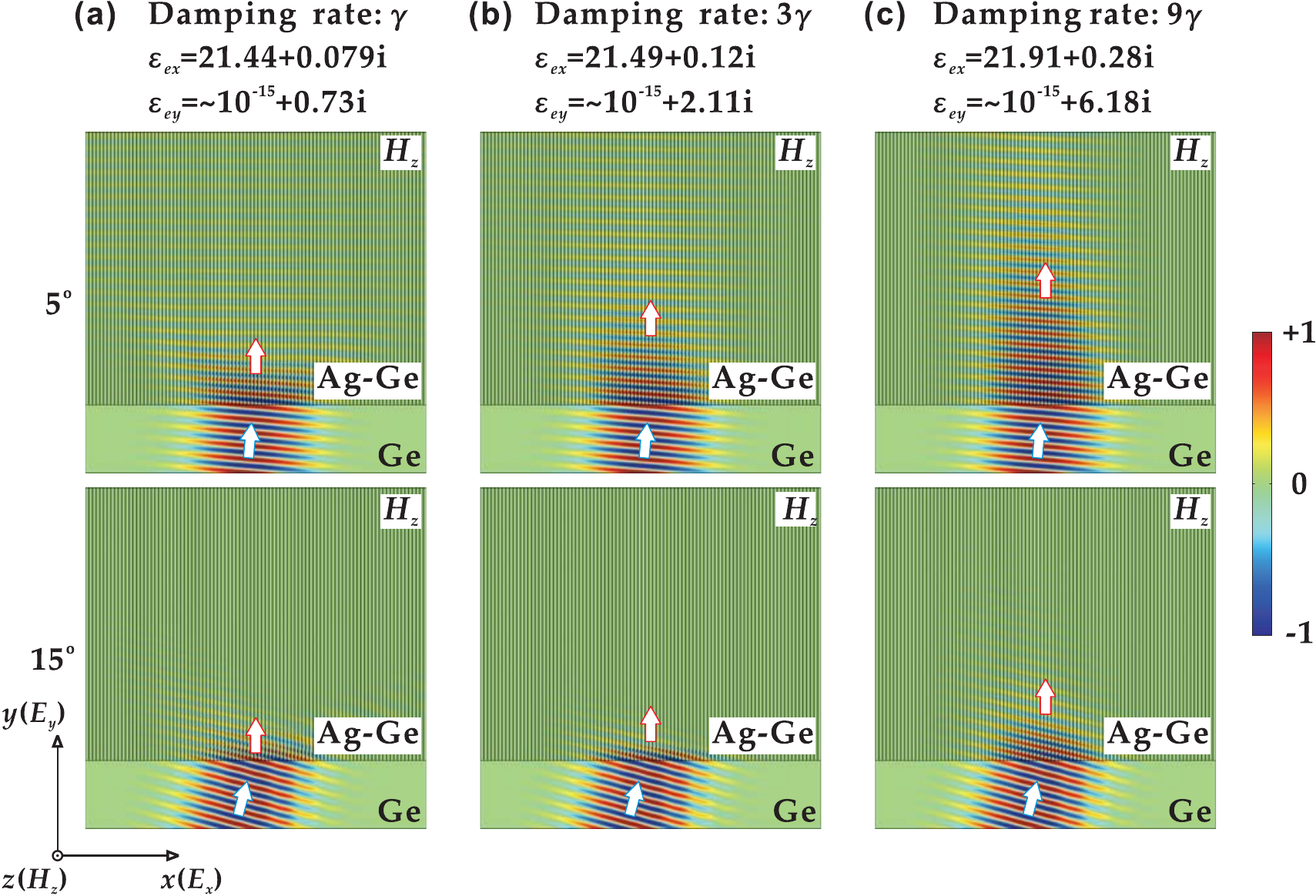}
    \caption{Lei Sun, Simin Feng, and Xiaodong Yang}
    \label{fig:fig2}
\end{figure}

%%%%%% SIMULATION %%%%%%%%%%%%%%%%%%%%%%%%%%%%%%%%%%%%%%%%%%%%%%%%%%%%%%%%%%
Based on the effective medium theory \cite{Bergman1992SSP}, we numerically demonstrate this exotic
electromagnetic phenomenon with realistic metal-dielectric multilayered structures,
as depicted in Fig.~\ref{fig:fig2}.
The anisotropic ENZ metamaterial is constructed as a multilayer consisting
of alternating thin layers of silver and germanium stacking in the $x$-direction.
In order to reduce the reflection at the interface, the multilayered stack is placed
on a germanium substrate.
The thickness of each sliver-germanium pair (along the $x$-direction) is $100\,\mathrm{nm}$,
and the effective permittivity of the stack can be represented as \cite{Bergman1992SSP}
\begin{equation}
\label{eq:permittivity}
\begin{aligned}
	\varepsilon_{ex}
		&= \left(f/\varepsilon_{\mathrm{Ag}}
			+(1-f)/\varepsilon_{\mathrm{Ge}}\right)^{-1}, \\
	\varepsilon_{ey}
		&= f\varepsilon_{\mathrm{Ag}}+(1-f)\varepsilon_{\mathrm{Ge}},
\end{aligned}
\end{equation}
where $f$ is the filling ratio of the silver.
The permittivity of the silver follows the simple Drude model
$\varepsilon_{\mathrm{Ag}}=\varepsilon_{\scriptscriptstyle\infty}
+\omega_{p}^{2}/(\omega(\omega+\ii\alpha\gamma))$ with permittivity constant
$\varepsilon_{\scriptscriptstyle\infty}=5.0$,
plasma frequency $\omega_{p}=1.38\times10^{16}\,\mathrm{rad}/\mathrm{s}$,
and damping rate of bulk silver $\gamma=5.07\times10^{13}\,\mathrm{rad}/\mathrm{s}$.
The additional factor $\alpha$ in the simple Drude model is used to
account for the increased loss due to surface scattering,
grain boundary effects in the thin silver film and inhomogeneous broadening
in the realistic experiment \cite{Yang2012NP,He2012OL}.
In the simulation, three different values of factor $\alpha$ are considered,
including $\alpha=1$, $\alpha=3$, and $\alpha=9$,
which results in three different damping rates: $\gamma$ (Fig.~\ref{fig:fig2}\blue{(a)}),
$3\gamma$ (Fig.~\ref{fig:fig2}\blue{(b)}), and $9\gamma$ (Fig.~\ref{fig:fig2}\blue{(c)}).
Moreover, the permittivity of germanium is $\varepsilon_{\mathrm{Ge}}=19.010+0.087\ii$ \cite{Adachi1988PRB},
and the ENZ frequency, at which $\mathrm{Re}(\varepsilon_{ey})=0$,
is designed at $193.4\,\mathrm{THz}$ ($1.55\,\mu\mathrm{m}$),
thus based on Eq.~\eqref{eq:permittivity} the filling ratio of the silver is
$f=0.129$ (Fig.~\ref{fig:fig2}\blue{(a)}),
$f=0.130$ (Fig.~\ref{fig:fig2}\blue{(b)})
and $f=0.145$ (Fig.~\ref{fig:fig2}\blue{(c)}), with respect to the different damping rates.
A TM polarized Gaussian beam at the ENZ frequency $193.4\,\mathrm{THz}$ is considered
in the simulation, and two angles of incidence, $5^{\circ}$ and $15^{\circ}$,
are calculated for each damping rate.
Since lower damping rate leads to smaller material loss $\mathrm{Im}(\varepsilon_{ey})$
along the $y$-direction in the multilayered stack, it is clear that the Gaussian beam
almost cannot couple into the silver-germanium multilayer stack and propagate inside
for both angles of incident, although there is some optical scattering along the interface
due to the layered structures.
Furthermore, it is worth mentioning that the Gaussian beam splitting phenomenon shown
in Fig.~\ref{fig:fig2}\blue{(a)} inside the multilayered stack is caused by the optical
nonlocality in ENZ metamaterials \cite{Pollard2009PRL,Orlov2011PRB}.
As the damping rate increases, as shown in Fig.~\ref{fig:fig2}\blue{(b)}
and \ref{fig:fig2}\blue{(c)}, more energy of the
Gaussian beam can enter into the stack for both angles of incidence,
especially for the lower angle of incidence, resulting in a clear demonstration
of the loss enhanced transmission in realistic metamaterial structures.
Moreover, the simulation results also indicate that the Gaussian beam propagates
inside the multilayered stack normal to the interface between the stack and germanium
substrate, which agrees well with the above analysis.

%%%%%% CONCLUSIONS %%%%%%%%%%%%%%%%%%%%%%%%%%%%%%%%%%%%%%%%%%%%%%%%%%%%%%
To conclude, we have verified the extraordinary transmission enhancement and beam collimation
caused by the material loss in anisotropic ENZ metamaterials through numerical
simulations, and explained this phenomenon by applying the IFCs of anisotropic
ENZ metamaterials.
Furthermore, we also revealed the possibility in experimental realization of this
phenomenon by using silver-germanium multilayered structures, which will stimulate
experimental efforts to fabricate such anisotropic ENZ metamaterials and characterize
this exotic optical phenomenon.

%%%%%% ACKNOWLEDGEMENT %%%%%%%%%%%%%%%%%%%%%%%%%%%%%%%%%%%%%%%%%%%%%%%%%%%%%%
This work was partially supported by the Department of Mechanical and Aerospace Engineering,
the Materials Research Center,
the Intelligent Systems Center,
and the Energy Research and Development Center at Missouri S\&T,
the University of Missouri Research Board,
and the Ralph E. Powe Junior Faculty Enhancement Award.
S. Feng was supported by NAVAIR's ILIR program.
The authors acknowledge Y. He for his useful discussions about this work.

%%%%%% REFERENCES %%%%%%%%%%%%%%%%%%%%%%%%%%%%%%%%%%%%%%%%%%%%%%%%%%%%%%%%%%%

%%%%%% FIGURES %%%%%%%%%%%%%%%%%%%%%%%%%%%%%%%%%%%%%%%%%%%%%%%%%%%%%%%%%%%
\clearpage
\newpage
\section*{Figure Captions}
\textbf{FIG.~1}. (Color online)
The IFCs of the anisotropic ENZ metamaterial with different material losses,
and the simulation results of the loss enhanced transmission in the anisotropic
ENZ metamaterials.
For the anisotropic ENZ metamaterial, the IFC of $k_{x}\sim\mathrm{Re}(k_{y})$ is
plotted as red solid curve, and the corresponding IFC of $k_{x}\sim\mathrm{Im}(k_{y})$
is plotted as red dashed curve.
The IFC of air is plotted as blue solid curve.
The Gaussian beam is TM polarized and the distribution of the real part of
$H_{z}$ component is plotted.
The incoming Gaussian beam is denoted by the wave vector $\vec{k}_{\mathrm{in}}$ and
the Poynting vector $\vec{S}_{\mathrm{in}}$, with different angles of incidence,
including $15^{\circ}$, $30^{\circ}$, $45^{\circ}$, and $60^{\circ}$.
The propagating Gaussian beam inside the anisotropic ENZ metamaterial is denoted
by the wave vector $\vec{k}_{\scriptscriptstyle\mathrm{ENZ}}$
and the Poynting vector $\vec{S}_{\scriptscriptstyle\mathrm{ENZ}}$.

\vspace{5.0mm}
\textbf{FIG.~2}. (Color online)
The simulation results of the loss enhanced transmission in realistic silver-germanium
multilayered stack, placed on a germanium substrate.
The shaded area represents the multilayer stacking in the $x$-direction.
The incoming Gaussian beam is TM polarized and the distribution of the real part of $H_{z}$ component
is shown at the ENZ frequency $193.4\,\mathrm{THz}$ ($1.55\,\mu\mathrm{m}$).
The propagation direction of the Gaussian beam is denoted by hollowed arrows.
Two different angles of incident, $5^{\circ}$ and $15^{\circ}$,
are considered with respect to three different damping rates,
including $\gamma$, $3\gamma$, and $9\gamma$.

%%%%%% FIGURES %%%%%%%%%%%%%%%%%%%%%%%%%%%%%%%%%%%%%%%%%%%%%%%%%%%%%%%%%%%
%%% FIGURE 1 %%%
\clearpage
\newpage

%%% FIGURE 2 %%%
\newpage

\end{document}